\documentclass[12pt]{article} 
\usepackage{amsmath} 
\usepackage{amssymb}
\usepackage{epsfig}
\usepackage{apacite}
\newtheorem{lemma}{Lemma}
\newtheorem{corollary}{Corollary}
\newtheorem{claim}{Claim}

\newcommand{\dom}{\operatorname{dom}} 
\newcommand{\pdom}{\operatorname{dom}_p} 
\newcommand{\pow}{\operatorname{pow}}

\newcommand{\cA}{\mathcal{A}}
\newcommand{\cB}{\mathcal{B}}

\newcommand{\cP}{\mathcal{P}}
\newcommand{\cR}{\mathcal{R}}
\newcommand{\cX}{\mathcal{X}}
\newcommand{\cY}{\mathcal{Y}}

\newcommand{\nat}{\: \natural \:}

\newcommand{\tm}{\tilde{\mu}}
\newcommand{\mq}{\mu_q}
\newcommand{\hm}{\hat{\mu}}
\newcommand{\hOm}{\hat{\Omega}}
\newcommand{\hSg}{\hat{\Sigma}}

\newcommand{\bA}{\Bar{A}}
\newcommand{\bB}{\Bar{B}}
\newcommand{\bC}{\Bar{C}}

\font\openface=msbm10 at10pt
\def\Minkowski     {{\hbox{\openface M}}}

\def\Reals         {{\hbox{\openface R}}}
\def\Complexes     {{\hbox{\openface C}}}

\def\ofS           {{\hbox{\openface S}}}

\begin{document}
\title {Comparing causality principles} 
\author{ 
Joe Henson\footnote{Department of 
Mathematics, University of California/San Diego, La Jolla,
CA 92093-0112, USA. E-mail: jhenson@math.ucsd.edu} 
} 
\maketitle
\begin{abstract}
The principle of common cause is discussed as a possible fundamental principle of physics.  Some revisions of Reichenbach's formulation of the principle are given, which lead to a version given by Bell.  Various similar forms are compared and some equivalence results proved.  The further problems of causality in a quantal system, and indeterministic causal structure, are addressed, with a view to defining a causality principle applicable to quantum gravity.\\

\end{abstract}
\vskip 1cm
\section{Introduction}

%
In the search for new and more general theories of nature, it is of interest to ask which physical principles will survive in the next fundamental theory, and which will be only approximately true.  Candidate answers have been of use in the formulation of theories in the past, and one might hope that they may be again, for instance in quantum gravity.  Most directly, given a kinematical framework, physical principles can be used to constrain the dynamics until only a small class of theories remain (an example being the derivation of general relativity from the principle of equivalence, general covariance, \textit{etc.})  In the causal set quantum gravity program, which is based on a simple kinematical structure, this approach is particularly natural, and has already been used to formulate a stochastic dynamics for causal sets \cite{CSG}.

As superluminal signalling (or more objectively, superluminal influence) is widely held to be impossible in current theories, and moreover seemingly impossible to square with relativity, a condition based on this would be a strong candidate for a fundamental principle \cite{Dowker:2004}.  What form would such a principle take?  It would be desirable to avoid two things: subjective statements involving observers, and controversies in the philosophy of causation.  Therefore, so far, the causal principles of most interest to physics have been those that give conditions, in terms of probabilities, that are meant to be physically reasonable without the need for one agreed definition of causes, effects and so on.  These conditions, and their names, are many and various; screening off, the Reichenbach principle of common cause (PCC) \cite{Reichenbach}, local causality \cite[pp. 52-66]{Bell}, and stochastic Einstein locality (SEL) \cite{Hellman} are most widely used.  Uffink's \citeyear{Uffink} provides a good introduction to the PCC, raises some of the questions that I attempt to deal with below, and criticises other forms of the principle.

The seeming variety of formulations might be taken as speaking against causality as a fundamental principle.  However, as the conditions are clarified, generalised and otherwise revised, they tend either to fall victim to paradox, or to converge to equivalence.  An example of this is given in the first section of this article, where Reichenbach's PCC becomes a statement resembling Bell's version of screening off after a few well motivated revisions (Butterfield's \citeyear{Butterfield} includes a related discussion of SEL).  In the same section, screening off is seen to be immune to certain paradoxes that afflict other PCC-like principles.  Various other forms of the principle are compared and some claims of equivalence proved.  An argument from \cite{Dowker:2004} is also touched on below: that even ``weak relativistic causality'', a very weak ``common cause'' condition, is equivalent to screening off, if taken to be true when probabilities are conditioned on past events.

All this would not be surprising - if quantum mechanics did not violate screening off.  As it is, all of the stochastic definitions of causality fall down here.  But still, quantum mechanics does not allow superluminal signalling, leading many to think that there really is no superluminal influence, \textit{i.e.} that some principle of relativistic causality should still hold here.  If this attitude is taken, then there is some assumption that has been used so far which needs to be dropped.  One such assumption is this: that the framework of stochastic processes is a sufficiently general one to describe our physical theories, containing all relevant information about the system in question, in particular everything that could possibly be relevant to causality.  Recent developments in quantum mechanics cast doubt on this.  Quantum mechanics can be described as \textit{quantum measure theory} (\citeauthor{QuantumMeasure}, \citeyearNP{QuantumMeasure}, \citeyearNP{QM2}; \citeauthor{Martin:2004xi}, \citeyearNP{Martin:2004xi}), a generalisation of probability measure theory, giving rise to the idea of a \textit{quantal process} as a more fundamental framework than the stochastic process. In \cite{Dowker:2004} a candidate principle called \textit{quantum screening off} is derived using this line of thought, and shown to be obeyed by local relativistic QFT.  In Section 3 below, quantum screening off is reviewed and two forms compared, in close analogy to the previous section's discussion of stochastic screening off.

If some kind of PCC can indeed survive quantum mechanics, there is another hurdle to jump before it could be applied to any quantum gravity theory: indeterministic causal structure.  Even in the stochastic framework, the problems involved in defining a causal principle for ``background independent'' theories are severe.  Reconciling causality with general covariance, and finally moving to the quantal process framework, present further difficulties.  These are expounded at the end of section 3.

\section{From Reichenbach to Bell}
\label{SecReichenbach}
\subsection{Reichenbach's principle and two generalisations}
The first attempt to formulate the PCC is given in terms of the probabilities of certain events.  Not surprisingly, then, we are helped considerably by a good definition of events.  If we call the space of all possible histories of the system in question $\Omega$, then an event is a set in an appropriate $\sigma$-algebra $\Sigma$ of subsets of $\Omega$.  All this means is that if $\{A_i\}$ is a countable set of events, then $\bigcup_i A_i$ and $A_i^c$, the complement of $A_i$, should also be events.  The thought behind this is that, if we have defined the events ``the temperature was below $x$ in Dublin'', then we must also be able to define events like ``The temperature was above $x$ and below $y$ in Dublin'', and so on.  This piece of formalism helps to solve one of the major problems with the PCC, ``Bernstein's paradox'', later on.  We also have the probability measure $\mu:\Sigma \longrightarrow \Reals$ obeying the usual axioms.  In this language, what might usually be called $P(A \& B)$ becomes $\mu(A \cap B)$, $P( \neg A)$ is $\mu(A^c)$, etc.  As usual, conditional probabilities are defined by $\mu(A|B)=\mu(A \cap B)/\mu(B)$.  A \text{partition} of the space $\Omega$ is an exhaustive list of alternative events, \textit{i.e.} a disjoint set of events $\Phi$ such that $\bigcup_{A \in \Phi} A = \Omega$.

A sensible definition of dependence of events for the purpose of defining the PCC is 
\begin{equation}
\label{cor}
\mu(A)\mu(B) \neq \mu(A \cap B).
\end{equation}
Eqn.(\ref{cor}) says that $\mu(B|A)\neq \mu(B)$, \textit{i.e.} that if we are given $A$, then that changes the probability we attribute to $B$ (and \textit{vice-versa}).
%

Consider the following situation.  Two illusionists, one in Athens and the other in Brussels, each toss a coin at the same time.  The event $A$ of the coin in Athens coming up heads is correlated to the event $B$ of the coin in Brussels coming up heads.  Are we astounded by this feat?  Not if the illusionists had previously met and randomly selected one of 2 pairs of biased coins to be used.  The correlation is nothing out of the ordinary if, after conditioning our probabilities on which pair of coins was selected, the correlation disappears.  For instance,
\begin{alignat}{3}
&\mu(A)=0.5 \text{, }& &\mu(B)=0.5 \text{, }& &\mu(A \cap B)=0, \\
&\mu(A|C)=1 \text{, }& &\mu(B|C)=0 \text{, }& &\mu(A \cap B|C)=0, \\
&\mu(A|C^c)=0 \text{, }& &\mu(B|C^c)=1 \text{, }& &\mu(A \cap B|C^c)=0, \\
&\mu(C)=0.5,
\end{alignat}
where $C$ is the event of selecting the first pair of coins.

Considering examples similar to this, Reichenbach proposed the following principle for positively correlated events.
\paragraph{The original PCC:} if $A$ and $B$ cannot be causally connected,
\footnote{In Reichenbach's definition, he referred to $A$ and $B$ as being simultaneous.  This class of events is very limited in non-relativistic physics and undefined in relativistic theories, and therefore not very useful.  The phrase ``cannot be causally connected'' is to be clarified below.}
 and $\mu(A \cup B) > \mu(A)\mu(B)$, then there exists an event $C$ such that
\begin{align}
\label{R1}
 \mu(A \cap B|C)      &= \mu(A|C)\mu(B|C), \\
\label{R2}
 \mu(A \cap B|C^c) &= \mu(A|C^c)\mu(B|C^c), \\
\label{ACAnotC}
 \mu(A|C)             &> \mu(A|C^c), \\
\label{BCBnotC}
 \mu(B|C)             &> \mu(B|C^c), 
\end{align}

As noted by Uffink \citeyear{Uffink}, conditions (\ref{ACAnotC},\ref{BCBnotC}), and the requirement of positive correlation are really just window dressing, expressing the idea that $C$ is a ``cause'', and that a cause is usually taken to mean something that makes an ``effect'' more likely to happen.  But if the aim of the PCC is to demand an explanation of certain correlations, the negative correlations need explaining too, and the explanation by (\ref{R1},\ref{R2}) works even if (\ref{ACAnotC},\ref{BCBnotC}) are not true, for instance in the above example.  Therefore $C$ does not have to be a ``cause'' in the sense that it makes $A$ and $B$ more likely, and this word will be avoided in favour of more neutral terms.
\paragraph{First revision of the PCC:}if $A$ and $B$ cannot be causally connected and (\ref{cor}) is true, then there exists an event $C$ such that (\ref{R1}) and (\ref{R2}) are satisfied.

There is another fairly obvious generalisation.  What if our illusionists were to choose between more than two pairs of coins?  This situation could well violate the first revision of the PCC, as now (\ref{R2}) may fail to hold.  But this is no more amazing than their first plan; there is nothing physically unreasonable about the correlation.  We need another principle.
\paragraph{Second revision of the PCC:} if $A$ and $B$ cannot be causally connected and (\ref{cor}) is true, then there is a partition $\Phi$ of $\Omega$ such that
\begin{equation}
\label{SO}
 \mu(A \cap B|C) = \mu(A|C)\mu(B|C) \quad \forall C \in \Phi. \\
\end{equation}
This definition is weaker than the previous one since $\{C,C^c\}$ is a partition of $\Omega$.  It has previously been proposed for different reasons \cite{Uffink}.

Motivated by this, a \textit{screening event} is defined as an event $C$ such that (\ref{cor}) and (\ref{R1}) hold for some $A$ and $B$; the definition is only a casual one and might be extended to events such that (\ref{cor}) and (\ref{R1}) are true when the probabilities are conditioned on some past event.
\subsection{Simpson's paradox and the necessity of a spatiotemporal causal structure}
What correlations are in need of explanation?  The vague formulation ``could not be causally connected'' needs clarification.  One answer is that events happen in spacetime, and spacetime has a causal structure.  (Care must be taken here with nomenclature: causal relations and causal structure will be used in the physicists' manner as describing the light-cone structure of spacetime, and are not directly related with any favourite philosophical notion of causation).  Events that occur in spacelike (\textit{i.e.} causally unrelated) regions should be the ones required to obey the PCC.  This gives a rigorous concept to replace ``could not be causally connected''.  For everyday use we might make an effective definition of causal structure: for instance, the coin tosses in the previous example could be considered as occurring in effectively causally unrelated regions.  There are all kinds of possible generalisations.  All that is required is a ``spatiotemporal'' partial order $\ofS$ in which events occur, \textit{e.g.} a weakly causal Lorentzian manifold
\footnote{A weakly causal Lorentzian manifold is one that contains no closed causal curves.}
 or a causal set.  More arguments will be presented in favour of this later on.

Using this definition requires some caution: it amounts to an acceptance that ``causal influences'' cannot propagate outside of the light-cone.  If the PCC were to fail using this definition of causal structure, it leaves the possibility that it is not Reichenbach's PCC, but the relativistic principle, that has failed --- a view embraced in pilot wave theory (``Bohmian mechanics'').

Subsets of $\ofS$, or ``regions'', will be denoted by calligraphic typeface.  As in relativity the past of a point $x \in \ofS$ (\textit{i.e.} the set of all points in $\ofS$ that are less than $x$ in the causal partial order, including $x$ itself) is called $J^-(x)$, and $J^-(\cX)=\bigcup_{x\in\cX} J^-(x)$.  The future set $J^+(\cX)$ is defined similarly.  The notation $\cX \nat \cY$ will be used to indicate that the regions $\cX$ and $\cY$ are spacelike, \textit{i.e.} that $J^-(\cX) \cap \cY =\emptyset$ and $J^-(\cY) \cap \cX =\emptyset$.
\subsubsection{Domains of decidability}
\label{SecDom}
Now the concept of the region in which an event occurs has become important.  For every event $A$, there is assumed to be a unique smallest region $\dom(A)$, such that knowing all the properties of the history in $\dom(A)$ enables us to decide (without further knowledge of $\mu$) whether $A$ occurred or not, called the \textit{least domain of decidability}.  This is not the only sensible prescription that could be made.  For instance, if the history space $\Omega$ was made up of solutions of the Maxwell equations, there could be several such regions which did not overlap.  If $A$ was a set of histories in a which beam of light passes through a region $\cA$ (a well enough defined event in the classical theory), then given the field in $\cA$ we could decide whether $A$ occurred or not.  But we could also do so given the field in a thickened spacelike slice of $J^-(\cA) \backslash \cA$.  However, the requirement that there be a single domain of decidability for each event is not restrictive.  The above situation could be easily represented by making the history space $\Omega$ the space of all possible configurations of the electromagnetic field, and simply setting the probability of the set of non-solutions to be 0.  In this larger history space, knowing all the properties of a history within one region $\cA$ decides no event that occurs outside $\cA$, without knowledge of $\mu$.

Again, some definitions will help.  It will not be necessary in this paper to construct any framework dealing with properties of histories, their relation to events, and their locations in the spacetime structure; it is enough to assign a least domain of decidability to each event by fiat, with some rules for consistency.   The least domain of decidability is defined as a function $\dom : \Sigma \longrightarrow \pow(\ofS)$ (where $\pow(\ofS)$ is the set of all subsets of $\ofS$) with the following properties:

\paragraph{} For all countable subsets $\Lambda$ of $\Sigma$,

\paragraph{(i)}   $\dom(X) \cap \dom(Y) = \emptyset \quad \forall X,Y \in \Lambda$ such that $X \neq Y$

$\Longrightarrow \; \dom(\bigcap_{X \in \Lambda} X) = \bigsqcup_{X \in \Lambda} \dom(X)$.

%
\paragraph{(ii)}  $\dom(X) = \dom(Y) \quad \forall X,Y \in \Lambda 
\Longrightarrow \dom(\bigcap_{X \in \Lambda} X) \subset \dom(Y) \quad \forall \: Y \in \Lambda$.
\paragraph{(iii)}  $\dom(X^c)=\dom(X)\ \quad \forall X \in \Sigma$.

\paragraph{(iv)} $\forall Z \in \Sigma$ s.t. $\dom(Z)=\cX \sqcup \cY$, $Z$ is a member of the $\sigma$-algebra generated by $\Gamma(\cX) \cup \Gamma(\cY)$, where $\Gamma(\cX)=\{X \in \Sigma :\dom(X) \subset \cX \}$.

\paragraph{} Here $\sqcup$ indicates disjoint union
\footnote{These provisional definitions may need modification.  It is not immediately clear that the countable unions in (i,ii) are strong enough.  For instance, if dom(``The scalar field holds value $x$ at point $\cX$'') is the point $\cX$, then dom(``The scalar field held value $x$ throughout region $\cR$'') should be $\cR$.  The question of whether this follows from (i-iv) is left unanswered for now as the definitions given are sufficient for the purposes of this article.}.  
Properties (i-iii) are fairly intuitive, while (iv) ensures some ``locality'' in the decidability of events:  it says that any statement about events in the region $\cX \cup \cY$ is a logical combination of statements about events in $\cX$ and $\cY$  (this assumption is considered in more detail later).  It will be said that ``$X$ occurs in $\cX$'' if $\dom(X) \subset \cX$.  One immediate consequence of (i) and (iii) is:

\paragraph{} For all countable subsets $\Lambda$ of $\Sigma$,

\paragraph{}   $\dom(X) \cap \dom(Y) = \emptyset \quad \forall X,Y \in \Lambda$ such that $X \neq Y$

$\Longrightarrow \; \dom(\bigcup_{X \in \Lambda} X) = \bigsqcup_{X \in \Lambda} \dom(X)$.
%
Similarly, (ii) and (iii) give:

\paragraph{} For all countable subsets $\Lambda$ of $\Sigma$,

\paragraph{}  $\dom(X) = \dom(Y) \quad \forall X,Y \in \Lambda 
\Longrightarrow \dom(\bigcup_{X \in \Lambda} X) \subset \dom(Y) \quad \forall \: Y \in \Lambda$.

It is also useful to note that $\dom(\emptyset) = \dom(\Omega)=\emptyset$ for any suitably nontrivial $\Sigma$ and $\dom$.  This can be seen by noting that, by (ii) and (iii), $\dom(\emptyset)= \dom(X^c \cap X) \subset \dom(X)$ for all events $X$, so as long as there are two events with disjoint domains of decidability, the only possibility is $\dom(\emptyset) = \emptyset$.  This weak assumption will be made throughout the following.

If some knowledge is gained about events in some region, what do we know about events in a subset of that region?  If $\cX \subset \cY=\dom(Y)$, then the \textit{restriction} $X=\Gamma_\cX(Y)$ of $Y$ to $\cX$ is defined as the intersection of all events $Z$ such that $Y \subset Z$ and $\dom(Z) \subset \cX$.  This restriction represents the most specific event decidable in $\cX$ given by $Y$.

A \textit{full specification} of a region $\cR$ is defined to be a non-empty event $F$ such that $\dom(F) \subset \cR$ and
\begin{equation}
 \dom(X) \subset \cR \Longrightarrow F \subset X \text{ or } F \subset X^c \quad \forall X \in \Sigma,
\end{equation}
\textit{i.e.} ``given $F$, all events that are decidable within $\cR$ are also decided''.  Let $\Phi(\cR)$ be the set of all full specifications of $\cR$.  The following lemmas are useful.

This lemma states that if everything that happens in two regions is known, then everything that happens in the union of these regions is known.
\begin{lemma}
\label{lema}
If $\cA$ and $\cB$ are disjoint regions and  $A$,$B$ are full specifications of them respectively, then $A \cap B$ is a full specification of $\cA \sqcup \cB$.
\end{lemma}
\paragraph{Proof:}Take an event $Z$ such that $\dom(Z) \subset \cA \sqcup \cB$.  From property (iv) of $\dom$, $Z$ is in the $\sigma$-algebra generated by $\Gamma(\cA) \cup \Gamma(\cB)$, $\Gamma(\cX)$ being defined as in the definition of property (iv).  From the definition of a full specification, we have that either $A \subset X$ or $A \subset X^c$ for all $X \in \Gamma(\cA)$, and either $B \subset Y$ or $B \subset Y^c$ for all $Y \in \Gamma(\cB)$.  It follows that $A \cap B \subset W$ or $A \cap B \subset W^c$ for all $W \in \Gamma(\cA) \cup \Gamma(\cB)$, and so, by the properties of a $\sigma$-algebra, $A \cap B \subset W$ or $A \cap B \subset W^c$ for all $W$ that are members of the $\sigma$-algebra generated by $\Gamma(\cA) \cup \Gamma(\cB)$.  This implies that $A \cap B \subset Z$ or $A \cap B \subset Z^c$.  So $\dom(Z) \subset \cA \cup \cB \Rightarrow A \cap B \subset Z \quad \forall Z \in \Sigma $, and the definition of a full specification is satisfied.
$\Box$

The next lemma states that if everything in a region is known, then everything in a subset of that region is known.
\begin{lemma}
\label{lemb}
Given a full specification $A$ of $\cA$ and a set $\cB \subset \cA$, $B=\Gamma_\cB(A)$ is a full specification of $\cB$.
\end{lemma}
\paragraph{Proof:}  Take $X$ such that $\dom(X) \subset \cB$.  From property (iii) of $\dom$ we also know that $\dom(X^c) \subset \cB$.  Since $\cB \subset \cA$, and $A$ is a full specification of $\cA$ this implies that either $A \subset X$ or $A \subset X^c$.  But from the definition of restriction, $B$ is a subset of all sets $Z$ such that $A \subset Z$ and $\dom(Z) \subset \cB$, so either $B \subset X$ or $B \subset X^c$.  This proves that $\dom(X) \subset \cB \Rightarrow B \subset X$ or $B \subset X^c \quad \forall X \in \Sigma $.
$\Box$
\begin{corollary}
\label{cora} if $\cR=\bigsqcup_i \cA_i$ for some finite set of regions $\cA_i$, then a full specification $F$ of $\cR$ can be written $F=\bigcap_i A_i$ where $A_i$ is a full specification of $\cA_i$.
\end{corollary}

\begin{lemma} $\Phi(\cR)$, for any region $\cR$, is a partition of $\Omega$.
\end{lemma}
\paragraph{Proof:} Every singleton set $f$ in $\Omega$ is a full specification of $\ofS$ because it has no subsets \footnote{It has been assumed here that $\Omega$ can be discribed as a union of singletons.}. By lemma \ref{lemb} every such $f$ is in a full specification of $\cR$, defined by $\Gamma_\cR(f)$.  No singleton can be in more than one full specification of $\cR$, so $\Phi(\cR)$ is a disjoint set.  $\Box$
  
\subsubsection{Avoiding Simpson's paradox: Bell's PCC}

The PCC as it stands looks weak enough; we have prevented the illusionists from fooling us into calling a physically reasonable correlation ``magic''.  But what if two wizards were aiming to do the opposite: fool us into believing that their magic was reasonable?  They could try a similar trick to the illusionists.  They are possessed of two magical pairs of coins which produce correlated outcomes with no common cause.  Before setting off to perform their magic, they randomly select one of the two pairs.  For instance (taken from Uffink \citeyear{Uffink}):
\begin{alignat}{3}
&\mu(A)=0.5 \text{, }& &\mu(B)=0.5 \text{, }& &\mu(A \cap B)=0.25, \\
&\mu(A|S)=0.5 \text{, }& &\mu(B|S)=0.5 \text{, }& &\mu(A \cap B|S)=0, \\
&\mu(A|S^c)=0.5 \text{, }& &\mu(B|S^c)=0.5 \text{, }& &\mu(A \cap B|S^c)=0.5,\\
&\mu(S)=0.5,
\end{alignat}
where $A$ and $B$ are as in the previous example, as $S$ is the event of choosing the first pair of magic coins.  This situation is not one that we would want our condition to allow.  Yet, without taking $S$ into account, it looks like there is no correlation, and the common cause principle is not violated.  This is known as Simpson's paradox.  An event like $S$ is called a \textit{Simpson event}: an event $S$ such that
\begin{gather}
\mu(A \cap B) = \mu(A)\mu(B), \quad
\mu(A \cap B|S) \neq \mu(A|S)\mu(B|S).
\end{gather}
It is in a sense the opposite of a screening event.

\begin{figure}[ht]
\centering \resizebox{4.5in}{1.5in}{\includegraphics{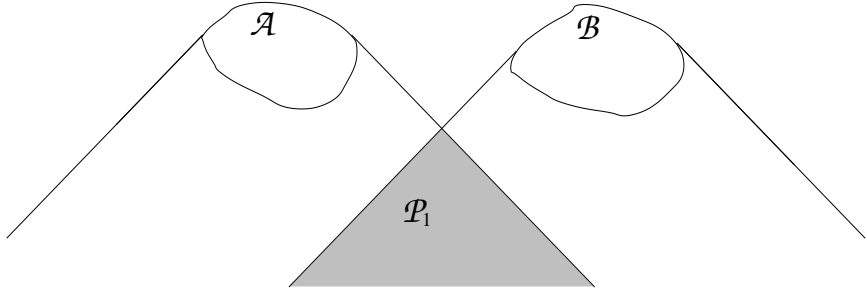}}
\caption{\small{A ``spacetime diagram'' of the regions involved in screening off (making $\ofS$ look like Minkowski space).  Two spacelike regions $\cA$ and $\cB$ are shown along with their past light cones.  $\cP_1$, shown in grey, is the mutual past of $\cA$ and $\cB$}.\label{screenfig}}
\end{figure}

The common cause principle as it stands looks too weak in the light of this.  It needs to be strengthened, so that it is still valid even if the probabilities are conditioned on events to the past of $A$ and $B$.  But what is meant by ``the past of $A$ and $B$'' without the explicit introduction of a causal structure?  If there is no such structure, then Simpson's paradox is hard to avoid, as Uffink \citeyear{Uffink} points out.  With the causal structure, a common cause principle can be formulated which seems well able to avoid the paradox.  The result is:

\paragraph{Screening Off (SO1):} For all events $A$ and $B$ with $\dom(A) \subset \cA$ and $\dom(B) \subset \cB$,  if $\cA \nat \cB$, then 
\begin{equation}
\label{SO1}
 \mu(A \cap B|C) = \mu(A|C)\mu(B|C) \quad \forall C \in \Phi(\cP_1), \\
\end{equation}
where $\cP_1$ is the mutual past $\cP_1=J^-(\cA) \cap J^-(\cB)$.  See fig.(\ref{screenfig}).

If the screening factors $C$ are full specifications of the past, then there is no room for a Simpson event in the past to reintroduce correlations.  Neither does the new definition seem too strong: we would expect, after conditioning on all events to the past, that two events in spacelike regions would be independent.  Thus through 3 strongly motivated modifications of Reichenbach's original principle we have reached some agreement with Bell's definition of causality \cite[pp. 52-66]{Bell} (in which he named a similar ``screening off'' principle ``local causality'').

Some justification should be given for requiring (\ref{SO1}) for any regions $\cA$ and $\cB$ such that $\dom(A) \subset \cA$ and $\dom(B) \subset \cB$, rather than only those such that $\dom(A) = \cA$ and $\dom(B) = \cB$.  This distinction is not as great as it might appear.  Consider a countable set of events $\{A_i\}$ such that $A= \bigsqcup_i A_i$, and let $\dom(A_i)=\cA$ for all $A_i$.  From properties (ii) and (iii) of $\dom$, the region $\dom(A)$ is contained in $\cA$ (see the immediate consequnces of the properties of $\dom$).  Requiring SO1, with this $\cA$, holds for each $A_i$ with $B$ implies that it must also be true for $A$:
\begin{gather}
 \mu(A_i \cap B|C) = \mu(A_i|C)\mu(B|C) \quad \forall C \in \Phi(\cP_1), \\
\Longrightarrow 
\mu( ( \, \bigsqcup_i A_i) \cap B|C) = \mu( (\, \bigsqcup_i A_i )|C)\mu(B|C) \quad \forall C \in \Phi(\cP_1),
\end{gather}
so the definition requiring that $\dom(A) = \cA$, when applied to all the events $A_i$, implies the definition requiring $\dom(A) \subset \cA$ in this case.  Such a case can be constructed for any $\cA \supset \dom(A)$ as long as there are events with $\cA$ as their least domain of decidability.

\subsubsection{Which past?}

No justification of the use of the mutual past has been given above.  The following principle seems just as well motivated:
\paragraph{A Second Screening Off (SO2):} For all events $A$ and $B$ with $\dom(A) \subset \cA$ and $\dom(B) \subset \cB$, if $\cA \nat \cB$, then
\begin{equation}
\label{SO2}
 \mu(A \cap B|C) = \mu(A|C)\mu(B|C) \quad \forall C \in \Phi(\cP_2), \\
\end{equation}
where $\cP_2$ is the \textit{joint} past $\cP_2=J^-(\cA) \cup J^-(\cB) \backslash (\cA \cup \cB)$.

Which of these is the most physically reasonable?  If SO1 is accepted, might Simpson events occur in $\cP_2$ but outside of $\cP_1$?  In fact, this cannot happen.  It is ruled out when SO1 is applied to pairs of events other than the pair $\{A,B\}$ in question.  The two definitions turn out to be equivalent, at least under a simplifying assumption.  The assumption is that there are only a finite number of full specifications of any region.

To see this we will need the following lemma in probability:
\begin{lemma}
\label{lemd}
If
\begin{gather}
\forall X \in \Phi(\cX), Y \in \Phi(\cY),  \notag \\
\label{ld1}
\mu(A \cap Y)\mu(B \cap X) = \mu(A \cap B \cap X \cap Y) \quad \text{, and} \\
\label{ld2}
\mu(A \cap Y)\mu(X) = \mu(A \cap X \cap Y) \quad \text{, and} \\
\label{ld3}
\mu(Y)\mu(B \cap X) = \mu(B \cap X \cap Y) \quad \text{, and} \\
\label{ld4}
\mu(Y)\mu(X) = \mu(X \cap Y) \
\end{gather}
then
\begin{equation}
\label{ld5}
\mu(A|X \cap Y)\mu(B|X \cap Y)=\mu(A \cap B|X \cap Y) \quad
\forall X \in \Phi(\cX), Y \in \Phi(\cY).
\end{equation}
\end{lemma}

{\bf Proof:}  (\ref{ld1}) and (\ref{ld4}) imply
\begin{equation}
\mu(A \cap Y)\mu(B \cap X)\mu(X)\mu(Y) = \mu(A \cap B \cap X \cap Y)\mu(X \cap Y) \quad
\forall X \in \Phi(\cX), Y \in \Phi(\cY),
\end{equation}
Substituting from (\ref{ld2},\ref{ld3}) into the LHS, we have
\begin{equation}
\mu(A \cap X \cap Y)\mu(B \cap X \cap Y) = \mu(A \cap B \cap X \cap Y)\mu(X \cap Y) \quad
\forall X \in \Phi(\cX), Y \in \Phi(\cY),
\end{equation}
which is equivalent to (\ref{ld5}).  $\Box$

\begin{claim}\label{claima}
 SO1 implies SO2.
\end{claim}
\paragraph{Proof:}
Assume SO1.  Take any pair of events $A$ and $B$, such that $\dom(A) \subset \cA$ and $\dom(B) \subset \cB$ and $\cA \nat \cB$.  As above, $\cP_1=J^-(\cA) \cap J^-(\cB)$ and $\cP_2=J^-(\cA) \cup J^-(\cB) \backslash (\cA \cup \cB)$, and it is also useful to define $\cX=(J^-(\cA) \backslash \cA)\, \backslash J^-(\cB)$ and $\cY=(J^-(\cB) \backslash \cB)\, \backslash J^-(\cA)$.  SO1 will be applied to other pairs of events in order to show that SO2 also holds.

Consider the pairs of events $\{(A \cap X),(B \cap Y)\}$, $\{(A \cap X),Y\}$, $\{A,(B \cap Y)\}$ and $\{X,Y\}$, where $X$ is a full specification of the region $\cX$, and $Y$ is a full specification of $\cY$.  To apply SO1 to these pairs we need to establish that the members of each pair occur in spacelike regions, and find the mutual past of those regions. Recall that a full specification $X \in \Phi(\cX)$ is defined so that $\dom(X) \subset \cX$, and since $\cA$ is disjoint to $\cX$, $\dom(A \cap X) \subset \cA \cup \cX$, for all $X\in \Phi(\cX)$, from property (i) of $\dom$ (similarly for $B$ and $Y$).  It is also the case that $\dom(X) \subset \cA \cup \cX$ and $\dom(Y) \subset \cB \cup \cY$.  From this we can see that for all four pairs of events, the two events occur in the pair of regions $\{(\cA \cup \cX),(\cB \cup \cY)\}$ respectively.  It can be seen that $(\cA \cup \cX) \nat (\cB \cup \cY)$, and that the mutual past for this pair of regions is $\cP_1$.  This is true for any causal structure (from its definition as a partial order), and is especially clear for Minkowski space, with reference to fig. \ref{screenfig}.  Thus SO1 can be applied four times, each time conditioning on a full specification $C$ of the past region $\cP_1$.  Using the notation $\tm(\cdot)=\mu(\cdot|C)$, SO1 gives:
\begin{gather}
\forall X \in \Phi(\cX), Y \in \Phi(\cY), C \in \Phi(\cP_1)  \notag \\
\tm(A \cap Y)\tm(B \cap X) = \tm(A \cap B \cap X \cap Y) \quad \text{, and} \\
\tm(A \cap Y)\tm(X) = \tm(A \cap X \cap Y) \quad \text{, and} \\
\tm(Y)\tm(B \cap X) = \tm(B \cap X \cap Y) \quad \text{, and} \\
\tm(Y)\tm(X) = \tm(X \cap Y) .
\end{gather}
From lemma \ref{lemd}, this implies that
\begin{equation}
\label{prf1}
\tm(A|X \cap Y)\tm(B|X \cap Y)=\tm(A \cap B|X \cap Y) \quad
\forall X \in \Phi(\cX), Y \in \Phi(\cY), C \in \Phi(\cP_1) . \\
\end{equation}
From corollary \ref{cora}, if $C \in \Phi(\cP_1)$, $X \in \Phi(\cX)$ and $Y \in \Phi(\cY)$, then $C \cap X \cap Y \in \Phi(\cP_2)$. Eqn. (\ref{prf1}) is therefore equivalent to (\ref{SO2}) and so SO2 is satisfied for the pair of events $\{A,B\}$.
$\Box$

Note that the assumption (that there is only a finite number of full specifications of any region) was needed here in order to use probabilities like $\mu(X)$ (which would always be 0 if the history space was continuous).  That is the only reason why it is needed in the proof of equivalence of SO1 and SO2.  As such the assumption could probably be removed, at the cost of making a more subtle use of measure theory in the proof of claim \ref{claima}.

\begin{claim}\label{claimb}
SO2 implies SO1.
\end{claim}

{\bf Proof:}  This claim is the easier to prove.  The argument is similar to one given by \citeauthor{Shimony} (\citeyearNP{Shimony}, p.165).  As above, take any pair of events $A$ and $B$, such that $\dom(A) \subset \cA$ and $\dom(B) \subset \cB$ and $\cA \nat \cB$.  We only need note that, if $\dom(A) \subset \cA$ and $\dom(B) \subset \cB$, then it is also true that $\dom(A) \subset \cA \cup \cX$ and $\dom(B) \subset \cB \cup \cY$, where $\cX$ and $\cY$ are defined as before.  The joint past of the pair of regions $\{\cA \cup \cX,\cB \cup \cY \}$ is $\cP_1$.  Since $\cA \cup \cX \nat \cB \cup \cY$, SO2 gives
\begin{equation}
 \mu(A \cap B|C) = \mu(A|C)\mu(B|C) \quad \forall C \in \Phi(\cP_1). \\
\end{equation}
From this, we know that, for any events $A$ and $B$ such that $\dom(A) \subset \cA$ and $\dom(B) \subset \cB$, if $\cA \nat \cB$, then the above equation holds.  This is exactly SO1.
$\Box$
\begin{corollary}
\label{corb}
SO1 is equivalent to SO2.
\end{corollary}

Generalisations of this result may be possible.  For example the following form of screening off is conjectured to be equivalent to SO1:

\paragraph{Generalised screening off:} For all events $A$ and $B$ with $\dom(A) \subset \cA$ and $\dom(B) \subset \cB$, if $\cA \nat \cB$, then
\begin{equation}
 \mu(A \cap B|C) = \mu(A|C)\mu(B|C) \quad \forall C \in \Phi(\cP'), \\
\end{equation}
where $\cP'$ \textit{contains} the mutual past $\cP_1=J^-(\cA) \cap J^-(\cB)$ and does not intersect the regions $J^+(\cA)$ or $J^+(\cB)$.  $\cP'$ could, for example, be $J^-(\cA) \backslash \cA$, as in Bell's original formulation.

\subsection{Avoiding Bernstein's paradox}

How does SO1 handle correlations between multiple events?  For some versions of the PCC, this is a fatal problem, which Uffink calls Bernstein's paradox \cite{Uffink}.  It stems from the observation that, even if $n$ events are each pairwise uncorrelated, it is not implied that there is no ``mutual correlation'', \textit{i.e.}
\begin{gather}
\mu(A_i \cap A_j)=\mu(A_i)\mu(A_j) \quad \forall i,j \in \{1,2,\dots,n\} \nRightarrow \\
\label{mutcor}
\mu(\bigcap_{i=1}^n A_i) = \prod_{i=1}^n \mu(A_i),
\end{gather}
where $\{A_i\}$ is a set of $n$ events.  What are we to do about such mutual correlations?  Do we need a further principle?  No - SO1 already covers this case.  Armed with the definition of events, we can see that, if all the events in $\{A_i\}$ occur in spacelike regions (\textit{i.e.} if $\dom(A_i) \nat \dom(A_j)$ for $i \neq j$), then $\{A_i,A_j\}$, $i\neq j$ are not the only spacelike pairs of events that are required to satisfy SO1.  From property (i) of $\dom$, it can be seen that for any $\Lambda \subset \{1,2,\dots,n\}$,
\begin{equation}
\dom(\bigcap_{i \in \Lambda} A_i)= \bigsqcup_{i \in \Lambda} \cA_i
\end{equation}
(where $\cA_i=\dom(A_i)$), so quite a few pairs events of the form $\bigcap_{i \in \Lambda} A_i$ occur in spacelike regions.  Assuming that \textit{all} spacelike pairs are independent, there is no mutual correlation:
\begin{align}
\mu(\bigcap_{i=1}^n A_i) &= \mu(A_1) \mu(\bigcap_{i=2}^n A_i) \\
                         &= \mu(A_1) \mu(A_2) \mu(\bigcap_{i=3}^n A_i)=...\\
                         &= \prod_{i=1}^n \mu(A_i)
\end{align}
Without presenting the argument in detail, it is claimed that considerations similar to those in the previous section show that the following is a consequence of SO1:
\paragraph{Screening off for multiple events:} Let $\{A_i\}$ be a set of $n$ events, and let $\dom(A_i) \subset \cA_i \; \forall i$, for some set of regions $\{\cA_i\}$.  If (\ref{mutcor}) is true and $\cA_i \nat \cA_j \quad \forall i,j \in\{1,\dots,n\}, \; i\neq j$, then 
\begin{equation}
\mu(\bigcap_{i=1}^n A_i|C) = \prod_{i=1}^n \mu(A_i|C) \quad \forall C \in \Phi(\cP_J), \\
\end{equation}
where $\cP_J$ is the joint past $\cP_J=\bigcup_{i=1}^n J^-(\cA_i) \backslash (\bigcup_{i=1}^n \cA_i)$, and $\cA_i=\dom(A_i)$.  This definition is similar to SO2; some other past regions can be substituted for $\cP_J$.  It is similar to a formulation by Uffink \citeyear{Uffink}.

\subsection{Further revisions?  Causality and locality}

SO1 is presented here as the most reasonable form of the common cause principle in stochastic theories with fixed causal structure.  But there are a few ways in which it could be modified for certain purposes.

At this point some terms need to be clarified.  What do the terms ``locality'' and ``causality'' stand for, and are they really the same thing?  Bell \citeyear{Bell} calls SO1 ``local causality''.  By ``local'' here he seems to be thinking of something like Einstein locality, a condition for deterministic theories which states that everything in a region $\cA$ should be fully determined by the configuration in $J^-(\cA) \backslash \cA$.  But other common uses of the word conflict with this in three ways.  (1) If a field theory is local in the sense that its action is an integral of a local function over spacetime, it is not necessarily Einstein local (consider electromagnetism with $c$ replaced by $2c$, for example);  (2) Einstein locality does not exclude the possibility that the occurrence of events in the ``deep past'' could directly influence those in the future (``temporal action at a distance''), since it does not state that the configuration in a region $\cA$ should be determined by the configuration in a thickened spacelike slice of $J^-(\cA) \backslash \cA$;  (3) Einstein locality does not require events to be defined locally.  Einstein locality can be thought of as a deterministic causality, but similar problems arise in stochastic and quantum theories. In all these ways, then, causality is not equivalent to locality.

The third of these differences has not been dealt with yet;  property (iv) of the least domain of decidability effectively requires that events be defined locally.  But this is not the only sensible definition that could be made.  In some cases it is not even physically realistic.  For example, in non-abelian gauge field theories, the gauge invariant ``observables'' cannot be defined locally.

Removing property (iv) would allow events to be non-local in general in the sense that the events that occur in a union of regions may not be deducible from those in the individual regions.  SO1 looks just as reasonable with this weaker definition of $dom$.  Some of the lemmas stated above would require modification if claims \ref{claima} and \ref{claimb} were to be proved in this case, however.  In particular, lemma \ref{lema} would no longer be true in general.

Another revision might also be considered.  There is a further possible definition of the past that has not hitherto been mentioned: \citeauthor{Penrose:1962} \citeyear{Penrose:1962} suggest that it should be any region $\cP_d$ that dissects $J^-(\cA) \cup J^-(\cB)$ into two parts, one containing $\cA$ and the other containing $\cB$.  This definition is stronger than SO1.  Is this strengthening justified?  Problems (2) and (3) above are relevant here.  If an event in the deep past can affect events in $\cA$, but no event in between, the Penrose-Percival principle may fail to take into account common causes for $A$ and $B$ lying in $\cP_1 \backslash \cP_d$.  Non-local events of the type already considered in this section could be in the past $\cP_1$ but not in every region $\cP_d$.  Thus this condition is stronger than SO1 in two ways.  If suitable locality conditions were assumed, such that both of these possibilities were ruled out, then it is a reasonable conjecture that Penrose and Percival's definition would also be implied by SO1 and SO2.

Finally, some modifications to the PCC might be desirable in the context of cosmology \cite[chap. 5]{Earman}.  Initial correlations are disallowed in SO2, \textit{i.e.} screening off is required even if $\cA$ or $\cB$ contain a part of the initial hypersurface.  There is no obvious reason to allow or disallow this, so the option remains open.  Consider the following definition.
\paragraph{Weakened screening off (SO2w):} For all events $A$ and $B$ with $\dom(A) \subset \cA$ and $\dom(B) \subset \cB$ \textit{such that} $\cA$ \textit{and} $\cB$ \textit{are of finite extent and do not contain any part of the initial hypersurface}, if $\cA \nat \cB$, then
\begin{equation}
 \mu(A \cap B|C) = \mu(A|C)\mu(B|C) \quad \forall C \in \Phi(\cP_2), \\
\end{equation}
where $\cP_2$ is the joint past $\cP_2=J^-(\cA) \cup J^-(\cB) \backslash (\cA \cup \cB)$.

Here the ``initial hypersurface'' should be taken to mean the set of points in the causal structure with no points to their past, and ``finite extent'' is intended to mean, in the case of a Lorentzian manifold, that there are no past-infinite causal curves in the region (no past-infinite chains in the case of a causal set, \textit{etc}.).  This definition stems from a suggestion of Rafael Sorkin's.  The proof that SO2 $\Rightarrow$ SO1 cannot be modified to show that SO2w $\Rightarrow$ SO1; at least, not without assuming that there are no correlations between events occurring on spacelike sections of the initial hypersurface (and that the causal structure is past-finite).  (It should be noted that, although this definition of causality may be weaker than that which is used to prove the Bell theorems, any attempt to evade them on this basis would require all EPRB experiments to be explained by initial spacelike correlations that are propagated toward the laboratory from far distant regions -- a conspiratorial scenario that is unbelievable for most scientists).  The question of whether the universe contained spacelike correlations in its earliest history is one for observation.  However, a definition of causality which treats initial points as special may not be natural for quantum gravity.

The big problems of quantum mechanics and dynamical causal structure also suggest revisions.  They are given their own section below.

\subsection{Other ways to justify SO1}
\label{wrc}
In \cite{Dowker:2004}, an easier way to reach the same conclusion about the PCC is given.
\paragraph{Weak relativistic causality:} if $\cA\nat\cB$, with $\cA \supset \dom(A)$ and $\cB \supset \dom(B)$, and $A$ is correlated to $B$, then there exists an event $C$ such that $\dom(C) \subset \cP_1=J^-(\cA) \cap J^-(\cB)$, and $A$ is correlated to $C$ and $B$ is correlated to $C$.

	This seems to be the weakest statement that could be made along the lines of a PCC.  However, in view of Simpson's paradox, this principle needs to be true even when the probabilities are conditioned on other events in $\cP_1$.  With this strengthening, weak relativistic causality can be shown to be equivalent to SO1.

It is also true that, if SO1 holds for some model, then it is possible to introduce an Einstein local deterministic hidden variable theory that reproduces the probabilities of that model \cite{Fine}
\footnote{The ``factoring'' condition used by \citeauthor{Fine} \citeyear{Fine} is actually not quite screening off, but a condition that follows from it when some assumptions of independence are made about the settings of experimental equipment (a ``freedom of choice'' assumption).}
.  The converse is also true.  (Incidentally, if the hidden variable theory is time reversal invariant, then there is also a corresponding principle of common \textit{effect}, which makes it impossible to define the arrow of time by the PCC, if such an underlying theory is assumed.)  This is another way to get back SO1 from a set of ``reasonable physical assumptions''; as discussed in the next section, it also highlights how reasonable and physical those assumptions really are.

\section{Quantum mechanics and dynamical causal structure}

It is well known that the above principle, SO1, is violated in standard relativistic quantum theory \cite[pp. 52-66]{Bell} (although arguably the corresponding experiment has not been successfully performed \cite{Percival}).  Correlations arise for which SO1 fails, even if we concoct ``hidden'' events in the past.  So presumably, if the illusionists were perseverant enough, they \textit{could} manage to produce a strange looking correlation, if we consider anything that breaks SO1 to be strange.  All the talk about what is ``physically reasonable'' seems to have been in vain; SO1 is not so physical after all.
%

There are different attitudes that can be taken to this situation, adapted here from the concise list of \citeauthor{Butterfield}'s \citeyear{Butterfield}: you can reject the whole notion of the PCC as a fundamental principle; you can see quantum mechanics as indicating that the PCC as set out above is inadequate and needs further revision; you can stick to a screening off principle, but reject relativistic causal structure and say that there is superluminal influence (as in Bohmian mechanics); or you might hope that future experimental results will be different from the predictions of quantum mechanics \cite{Percival}.  The following discussion takes the second tack.  Firstly, this is because superluminal signalling is still ruled out in quantum theory, and the PCC could reasonably be equated with relativistic causality, when the latter is defined appropriately.  Secondly, although screening off fails, there is still a vague sense in which the correlations arose as a result of some ``cause'' in the mutual past of the correlated events (for instance, in the classic example,  an entanglement is set up between two photons which might be seen as the ``cause'' of the correlation), and so it is reasonable to attempt to formalise this notion.  But, had quantum mechanics not come along, it is doubtful that anyone would have foreseen the need for a generalisation of SO1.  So what has gone wrong?

\subsection{Quantum screening off}

First, a decision must be made about how to view causality in quantum mechanics.  It is usual to claim that superluminal signalling is banned in standard QFT, and this is usually held to follow from the commutativity of spacelike operators.  This does guarantee that some events, such as the performance of a von Neumann measurement, will not be noticeable to a conspirator in a spacelike laboratory, in the absence of other measurements by third parties.  But this is not the end of the story.  This kind of discussion of superluminal signalling is based on external agents carrying out measurements and other operations on a quantum field, in attempts to send spacelike signals.  Even apart from the fact that naively applying this framework to QFT can be shown to \textit{allow} superluminal signalling \cite{Sorkin:1993gg}, it seems inappropriate for cosmological theories.  More preferable to this would be a definition of causality that is given in terms of the dynamics of the quantum system, without reference to classical observers of any kind.  Then the search for a ``relativistic causality'' principle becomes more like the discussion of the PCC that is given above for the stochastic case.  For more of this argument see \cite{Dowker:2004}.

If an observer independent, quantum PCC is being sought, then some idea needs to be given of how to understand quantum mechanics at all without observers.  Most appropriate for the present purpose, since we have already begun to think about histories, are the various ``histories'' approaches.  The details of the interpretations will not be needed here; the formulation of quantum mechanics associated with them is more relevant.  The central idea is the replacement of the stochastic measure $\mu:\Sigma\longrightarrow \Reals$ used above with a non-additive \textit{quantal measure} $\mq:\Sigma\longrightarrow \Reals$, obeying generalised axioms (for a full introduction to this idea see \citeNP{QuantumMeasure}) .  Equivalent to this, and sometimes easier to work with, is the decoherence functional $D:\Sigma\times \Sigma \longrightarrow \Complexes$ \cite{Hartle:1992as}  which has the following properties:
\paragraph{Hermiticity:} $D(A,B)=D^*(B,A)$ for all $A,B \in \Sigma$
\paragraph{Positivity:} $D(A,A) \geq 0$ for all $A \in \Sigma$
\paragraph{Additivity:} $D(A \sqcup B, C)= D(A,C)+D(B,C)$ for all $A,B \in \Sigma$
such that $A$ and $B$ are disjoint.  This can be extended to countable sums if necessary.
\paragraph{Normalisation:} $D(\Omega,\Omega)=1$ .

The value of the quantal measure is defined as $\mq(A)=D(A,A)$.  A quantal process is defined by the triple $\{ \mq,\Sigma,\Omega \}$, and all the definitions of events and least domains of decidability may be carried over
\footnote{This procedure, of defining events only on $\Sigma$ and not on some ortholattice as in \cite{Isham:1995vt}, is equivalent to picking a preferred basis (\textit{e.g.} position of particles) in which to define the histories.  This is still very much a quantum process, as pairwise interference of alternatives, which may be taken to be the defining quality of a quantal process, is still present.}.

Now there is a potential answer to the question of why SO1 is too strong for quantum mechanical systems:  quantum mechanics is not naturally described as a stochastic process.  But all of the reasoning in section \ref{SecReichenbach} was based on the assumption that a stochastic process defined by the triple $\{ \mu,\Sigma,\Omega \}$ can indeed adequately capture the dynamics of the universe --- in particular, that it can contain all information relevant to a principle of common cause.  Therefore we need to drop this assumption and work with a quantal process instead of a stochastic process.

Since the interpretation of the quantum measure is not so straightforward as that of the probability measure, what is ``physically reasonable'' and what is not becomes shadier here, making arguments for a quantum version of screening off less compelling.  It is significant that for some events, the fact of whether they occurred or not can never be known.  Should such events be allowed as the analogues of screening events here (can they be ``causes'')?  Should the PCC still hold when they are conditioned on?  Taking a realistic approach, the most natural (and fruitful) answer seems to be yes
\footnote{In the decoherence/consistent histories interpretation, the only ``knowable'' events are those that decohere.  If only these were considered when formulating a causality principle, we would not be any better off than we were in the stochastic case; the measure restricted to decohering events is a probability measure.}.

With suitable definitions of independence of events, and conditioning, a similar argument to one mentioned above in section \ref{wrc} can be used to derive a quantum version of screening off.  The following result is obtained in \cite{Dowker:2004}.
\paragraph{Quantum screening off (QSO):} Let $A,\bA$ and $B,\bB$ be events and $\cA \supset \dom(A)$, $\cA \supset \dom(\bA)$, $\cB \supset \dom(B)$ and $\cB \supset \dom(\bB)$.  If $\cA \nat \cB$, then 
\begin{equation}
 D(A \cap B \cap C, \bA \cap \bB \cap \bC)D(C,\bC) = 
 D(A \cap C, \bA \cap \bC)D(B \cap C, \bB \cap \bC) 
 \quad \forall C,\bC \in \Phi(\cP_1), \\
\end{equation}
where $\cP_1$ is the mutual past $\cP_1=J^-(\cA) \cap J^-(\cB)$.

This causality principle is formally true for local, relativistic QFT \cite{Dowker:2004}, providing good evidence that it is not too strong to be ``physically reasonable'', and reduces to SO1 when all events decohere (\textit{i.e.} when the decoherence functional is diagonal).

\subsubsection{Which past for QSO?}
\label{SecQWhich}

Conditioning on the joint or mutual past (the analogue of the choice between SO1 and SO2) is also equivalent here.  A cheap way to prove this is by a reformulation of the condition, making it possible to reuse the proof of the stochastic result.  The decoherence functional ranges over pairs of sets in $\Omega$.  Thinking of a measure space $\hOm=\Omega \times \Omega$, the decoherence functional can equivalently be represented as a function $\hm$ from a set of subsets $\hSg$ of $\hOm$ to the complex numbers.  Sets in $\hSg$, which will be called ``pseudo-events'', are defined as all sets $X \times Y$ where $X,Y \in \Sigma$, \textit{i.e.} as pairs of events.  From the definition of the decoherence functional, the function $\hm$ has some properties in common with a probability measure: it is additive, so that $\hm(A \sqcup B)=\hm(A)+\hm(B)$ for disjoint pseudo-events $A$ and $B$, and normalised to 1.  However, it is not a probability measure because it is not bounded above or below.  The least domain of decidability of a pseudo-event is defined by $\pdom(A \times \bA)=\dom(A) \cup \dom(\bA)$ (the definition for pseudo-events need not have properties (i-iv); it is shown in the appendix that this function does have the necessary properties to prove useful results).  A pseudo-event is called a full specification of $\cR$ iff it is the product of two events that are full specifications of $\cR$, and $\Phi_p(\cR)$ is defined accordingly as the set of all full specification pseudo-events (this is clearly a partition of $\hOm$).  In this framework, The QSO condition looks much like SO1:

\paragraph{A restatement of Quantum screening off (QSO1):} For all pseudo-events $A$ and $B$ with $\pdom(A) \subset \cA$ and $\pdom(B) \subset \cB$,  if $\cA \nat \cB$, then 
\begin{equation}
\label{QSO1}
 \hm(A \cap B \cap C)\hm(C) = \hm(A \cap C)\hm(B \cap C) \quad \forall C \in \Phi_p(\cP_1), \\
\end{equation}
where $\cP_1$ is the mutual past $\cP_1=J^-(\cA) \cap J^-(\cB)$.

There is a joint past alternative to this:

\paragraph{Another form of Quantum screening off (QSO2):} For all pseudo-events $A$ and $B$ with $\pdom(A) \subset \cA$ and $\pdom(B) \subset \cB$,  if $\cA \nat \cB$, then 
\begin{equation}
\label{QSO2}
 \hm(A \cap B \cap C)\hm(C) = \hm(A \cap C)\hm(B \cap C) \quad \forall C \in \Phi_p(\cP_1), \\
\end{equation}
where $\cP_1$ is the \textit{joint} past $\cP_1=J^-(\cA) \cup J^-(\cB) \backslash (\cA \cup \cB)$.

The proof of $QSO1 \Leftrightarrow QSO2$ is almost identical to that of $SO1 \Leftrightarrow SO2$.  The properties of the function $\hm$, and of the domain of decidability of pseudo-events, are similar enough to the properties of $\mu$ and the domain of decidability of events to ensure this.  However, since $\hm$ can be negative, it is now possible for $\hm(B)$ to be zero while $\hm(A \cap B)$ is non-zero, and so $\hm(A|B)$, the analogue of $\mu(A|B)$, is sometimes not well defined.  Therefore, for completeness, the proof is repeated for the quantum case, without use of such conditional statements, in the appendix.

\subsection{Dynamical causal structure}
QSO1 (SO1) is a formal statement of the notion that causal influences should propagate only within the light-cone, with the assumption that the quantal (stochastic) process is an adequate framework for our dynamical theory.  But in general relativity, and presumably in a successful quantum gravity, the light-cone structure of spacetime is itself dynamical.  How can a PCC be formulated if the regions in which events occur have different causal relations in different histories - or worse still, cannot be identified as being the same region in different histories?  If we also seek to impose the general covariance of GR on our theory, how does this affect these questions?  At the present time this seems to be the greatest problem for the definition of a PCC that could be of use in quantum gravity.  No solution is proposed below, but the problems are pointed out and some existing ideas are discussed.
\subsubsection{The stochastic case}
Before looking at any quantum indeterminism of causal structure, many of the significant problems with indeterministic causal structure can be brought to light by considering stochastic processes.  Previously, the least domain of decidability of an event was defined by fiat on a causal structure $\ofS$.  A deeper description would come from considering each history (\textit{i.e.} point in history space) as a list of values of properties attached to regions in the causal structure, like the values of a field on Minkowski space $\Minkowski$.  We are now moving to a theory in which the histories \textit{contain} the causal structure, which can be different in each history (for example, each history could be a different Lorentzian manifold).  The idea of domain of decidability is lost, as regions now have no significance for the whole history space.  Is it possible to restore enough of the idea to write down a condition similar to SO1 or SO2?

In SO2, probabilities are conditioned on a full specification of the past set
\footnote{A \textit{past set} in causal partial order is defined to be one that contains its own past: a region $\cR$ such that $\cR = J^-(\cR)$.}
 $\cP_2$.  Can this be done when causal structure is stochastic?  Firstly, any full specification $C$ of $\cP_2$ would still be an event, a set of histories in $\Sigma$.  It would be necessary to identify regions in different histories as being ``the same region'': a region like $\cP_2$ must be be defined on a set of histories containing $C$.  This is where the major conceptual problems lie.  If we knew how to do this, it would be possible to similarly define past sets $\cP_2 \cup A$ and $\cP_2 \cup B$, and make sure that the regions $A$ and $B$ were spacelike to each other.  With this done, a principle like SO2 could be recovered.  So the crucial question is:  when are two past sets in different histories \textit{the same} region?

A special strategy can be used to answer this when the history space consists of causal structures in which the points are uniquely labelled in some way (\textit{i.e.}, each point in the structure has a property that is shared by no other point in that structure, like a set of real numbers).  Two past sets in different histories are considered as being in the same region if they have the same causal structure, \textit{and} the same labelling of all points in that structure.  This is the general idea followed by \citeauthor{CSG} \citeyear{CSG} to define a causal set dynamics. A causal condition similar to SO2 (called ``labelled Bell causality'') is used to constrain a stochastic process on causal sets (formalised by \citeauthor{Brightwell:2002vw} \citeyear{Brightwell:2002vw}), leading to the ``classical sequential growth'' (CSG) model, which is to be understood as preparatory work for a full quantum dynamics.

The problem is that labelling like this threatens the principle of general covariance from GR: such ``co-ordinatisations'' are considered to be unphysical.  In the causal set dynamics \cite{CSG,Brightwell:2002vw}, general covariance is introduced as a fundamental principle, the dynamics being defined so that the labelling drops out of consideration.  But the Bell causality condition is still defined in terms of the labels.  What is the physical significance of labelled Bell causality in a theory that denies labels physical significance?  This riddle has not been fully answered, and as such work remains to be done on the motivational foundations of the CSG model.  As a start, this vague conjecture could be explored: that the general covariance and labelled Bell causality conditions of CSG imply some kind of generally covariant causality condition (as yet undefined).

It would be more satisfying to have a causality condition that did not rely on labelling.  Why not just identify a past set in different histories as the same region $\cP_2$ if it has the same causal structure?  One problem is that there may be more than one past set in a history that has the causal structure associated with $\cP_2$.  How are we to know which one is ``the same region'' as an isomorphic past set in another history?  To get rid of this problem, it could be assumed that there is only one copy of the past set.  A full specification of a region $\cP_2$ could be defined as consisting of histories containing only one past set that is isomorphic to a certain causal structure (that past set being identified in all such histories as the region $\cP_2$).

A number of questions arise.  Would such a statement be strong enough?  How can we know that there is no other past set in the universe identical to our own past?  Is it even reasonable to identify two past sets from different histories at all?  Or is it somehow justifiable to neglect these concerns?  At this (as yet) ragged edge of the concept of causality, the questions are still a little vague, and await better statements and answers.
\subsubsection{Dynamical causal structure \textit{and} quantum mechanics}

Moving from a stochastic to a quantal process, the problem of dynamical causal structure becomes even more virulent.  In the stochastic case, we have the problem of defining a region $\cP_2$ and a concept to replace a full specification in that region.  Once this is done, spacelike regions $\cA$ and $\cB$ can be given some meaning and SO2 resuscitated \cite<see >[ for a concrete example]{CSG}.  In the definition of QSO, however, we need \textit{two} full specifications of the same region $\cP_2$: $C$ and $\bC$.  If the causal structure in the two cases is different, how could this be made to make sense?  How could we make sure that the events $A$ and $\bA$ were both defined in the same region, spacelike to one in which $B$ and $\bB$ were defined?  Could some trick with labelled causal structures work here too?  These questions remain to be properly explored.

\section{Conclusion}

We have seen how, starting with Reichenbach's principle of common cause, Bell's causality principle, here termed screening off (SO), can be reached by a series of strongly motivated revisions.  Screening off roughly states that two events that occur in spacelike regions $\cA$ and $\cB$ must be independent when probabilities are conditioned on all events to the past of these two regions.  It has been shown that, modulo certain concerns of measure theory and locality, conditioning on the joint past in SO gives an equivalent definition to conditioning on the mutual past.  It has also been conjectured that \textit{further} conditioning on any other region spacelike to $\cA$ and $\cB$ would result in another equivalent definition.  With the assumption of stronger locality conditions, the Penrose-Percival causality condition is also conjectured to be equivalent.

	Some thoughts on quantum mechanics and the problem it poses for the PCC have been given, and the condition of quantum screening off \cite<from>{Dowker:2004} suggested as a solution. This causality condition is also seen to have two equivalent forms, one involving the mutual past and the other the joint past of two regions.  After this, the problem of indeterministic causal structure was addressed, although no concrete conclusions have been drawn on this subject as yet.

Apart from the conjectures already mentioned, there are many open questions relating to the use of the PCC as a fundamental principle of physics.  As a beginning, it would be satisfying to improve the results given in this article, extending the framework to non-local events by dropping property (iv) of the least domain of dependence, and getting rid of the requirement that there be a finite number of full specifications for any given region in the proof of corollary \ref{corb}.  Also, the discussion has only made formal use of histories as points in the history space $\Omega$; it would be helpful to make use of the idea of a history as a set of properties on the causal structure.  The concept of least domain of decidability could then be expanded so that it directly related to properties of histories instead of being introduced by hand.  This might be of some use in the definition of the PCC for theories with indeterministic causal structure.

Another closely related concept is that of stochastic Einstein locality (SEL) \cite{Hellman} \cite{Butterfield}.  Some formulations of this causality condition \cite<particularly SEL2 in>[which includes some discussion relating to the Reichenbachian PCC]{Butterfield} could be equivalent to the SO1 condition.  Indeed, SO1 is taken fairly directly from Bell's local causality \cite[pp. 52-66]{Bell}, and Hellman \citeyear{Hellman} cites Bell's work as the basis for his definition of SEL.  It would be interesting to see if a link between the framework discussed above and that of SEL could be made, and some theorems of equivalence proved.

The relationship between the SO1 condition and deterministic local hidden variables, only touched upon above, has significance not only with respect to the main motivations of this article, but also to more philosophical uses of the PCC, such as a definition of the direction of time.  Related to this is the fact that, if we have a theory which violates SO1, it may just be because we have failed to introduce the necessary common causes into our history space.  The idea of an \textit{extension} of a history space to include new events has not been dealt with in this article.  Startlingly, analogues of Fine's results for stochastic processes \cite{Fine} are suggested by new work on the quantal process.  Some of the new results \cite{Craig:2004} can be taken as examples of a conjecture: that, just as a SO1 obeying stochastic process can be consistently extended so that the history space contains only solutions of a deterministic causal (``Einstein local'') theory, a quantal process obeying QSO1 can be similarly extended
\footnote{More explicitly, a quantal process obeying QSO can be extended, so that all histories in the history space are solutions to a deterministic Einstein local dynamical law.}
.  This would give a puzzling new role to local hidden variable theories in quantum mechanics, without challenging the Bell theorems.  As Einstein locality is understood for deterministic, dynamical spacetimes in GR, it is possible that these observations could help to define the PCC in stochastic and quantal theories with indeterministic causal structure.  This subject will be investigated in future work.
\subsection*{Acknowledgements}
I thank Jeremy Butterfield for providing some of the literature on which this work was based, and for some helpful comments.  I have benefited greatly from discussions with Graham Brightwell, Raquel Garcia, Chris Isham and David Meyer, and from the conversations and collaborations with Fay Dowker and Rafael Sorkin that brought these issues to light.  Thanks are also due to Peter Morgan for some useful correspondence that led to a number of improvements in the article. This research was supported by DARPA grant F49620-02-C-0010.

\section*{Appendix:  Proof of QSO1 $\Leftrightarrow$ QSO2}

Firstly, the definition of the least domain of decidability for pseudo events, $\pdom(A_1 \times A_2)=\dom(A_1) \cup \dom(A_2)$ must be examined.  The following is to be compared to property (i) of $\dom$.

\begin{lemma} For all $X,Y \in \hSg$,
\label{pseudoi}

$\pdom(X) \cap \pdom(Y) = \emptyset$
$\Longrightarrow \; \pdom(X \cap Y) \subset \pdom(X) \sqcup \pdom(Y)$.
\end{lemma}
{\bf Proof:} Assume $\pdom(X) \cap \pdom(Y) = \emptyset$ for two pseudo-events $X$ and $Y$.  Let $X=X_a \times X_b$ and $Y=Y_a \times Y_b$.  From the definition of $\pdom$ we have that $\dom(X_i) \subset \pdom(X)$ and $\dom(Y_i) \subset \pdom(Y)$ for $i \in \{a,b\}$.  From property (i) of $\dom$ this gives $\dom(X_i \cap Y_j) \subset \dom(X) \sqcup \dom(Y)$ for $i,j \in \{a,b\}$.  This implies that $\pdom(X \cap Y) \subset \pdom(X) \sqcup \pdom(Y)$.  
$\Box$

An analogue of corollary \ref{cora} will also be needed.
\begin{lemma}
\label{pcora} if $\cR=\bigsqcup_i \cA_i$ for some finite set of regions $\{ \cA_i \}$, then a pseudo-event full specification $F$ of $\cR$ can be written $F=\bigcap_i A_i$ where $A_i$ is a pseudo-event full specification of $\cA_i$.
\end{lemma}
{\bf Proof:}  As stated in the main text, a pseudo-event like $F$ is a full specification of $\cR$ iff it is a product of two event full specifications of that region. In other words, $F=G \times H$, where $G$ and $H$ are full specifications of $\cR$.  From corollary \ref{cora} we have that $G=\bigcap_i B_i$ where $B_i$ is a full specification of the region $\cA_i$, and that $H=\bigcap_i C_i$ where $C_i$ is also a full specification of the region $\cA_i$.  This means that
\begin{equation}
F=\bigcap_i B_i \times \bigcap_jC_j= \bigcap_i A_i,
\end{equation}
where $A_i = B_i \times C_i$.  Since $A_{i}$ is a pseudo-event full specification of $\cA_i$, this proves the lemma.
$\Box$
 
The following is the quantal analogue of lemma \ref{lemd}.
 
\begin{lemma}
\label{Alemd}
If, for some pseudo-event $P_1$,
\begin{gather}
\forall X \in \Phi_p(\cX), Y \in \Phi_p(\cY), \notag \\
\label{Ald1}
\hm(A \cap Y \cap P_1)\hm(B \cap X \cap P_1) = \hm(A \cap B \cap X \cap Y \cap P_1) \hm(P_1) \quad \text{, and} \\
\label{Ald2}
\hm(A \cap Y \cap P_1)\hm(X \cap P_1) = \hm(A \cap X \cap Y \cap P_1) \hm(P_1) \quad \text{, and} \\
\label{Ald3}
\hm(Y \cap P_1)\hm(B \cap X \cap P_1) = \hm(B \cap X \cap Y \cap P_1) \hm(P_1) \quad \text{, and} \\
\label{Ald4}
\hm(Y \cap P_1)\hm(X \cap P_1) = \hm(X \cap Y \cap P_1) \hm(P_1) \
\end{gather}
then
\begin{gather}
\forall X \in \Phi_p(\cX), Y \in \Phi_p(\cY), \notag \\
\label{Bug} 
\hm(A \cap X \cap Y \cap P_1)\hm(B \cap X \cap Y \cap P_1)=\hm(A \cap B \cap X \cap Y \cap P_1) \hm(X \cap Y \cap P_1).
\end{gather}
\end{lemma}

{\bf Proof:}  Substituting (\ref{Ald1}) into (\ref{Ald4}) gives
\begin{multline*}
\hm(A \cap Y \cap P_1)\hm(B \cap X \cap P_1)\hm(X \cap P_1)\hm(Y \cap P_1) = \\
\hm(A \cap B \cap X \cap Y \cap P_1)\hm(X \cap Y \cap P_1) \hm(P_1)^2
\end{multline*}
\begin{equation}
\forall X \in \Phi_p(\cX), Y \in \Phi_p(\cY),
\end{equation}
Substituting from (\ref{Ald2},\ref{Ald3}) into the LHS, we have
\begin{gather}
\forall X \in \Phi_p(\cX \cap P_1), Y \in \Phi_p(\cY \cap P_1), \notag \\
\hm(A \cap X \cap Y \cap P_1)\hm(B \cap X \cap Y \cap P_1) = \hm(A \cap B \cap X \cap Y \cap P_1)\hm(X \cap Y \cap P_1),
\end{gather}
which is equivalent to (\ref{Bug}).
  $\Box$

\begin{claim}\label{Aclaima}
 QSO1 implies QSO2.
\end{claim}
\paragraph{Proof:}
Assume QSO1.  Take any pair of pseudo-events $A$ and $B$, such that $\pdom(A) \subset \cA$ and $\pdom(B) \subset \cB$ and $\cA \nat \cB$.  The regions $\cP_1$, $\cP_2$, $\cX$ and $\cY$ are defined as in the proof of claim \ref{claima}.

Consider the pairs of events $\{(A \cap X),(B \cap Y)\}$, $\{(A \cap X),Y\}$, $\{A,(B \cap Y)\}$ and $\{X,Y\}$, where $X$ is a pseudo-event full specification of the region $\cX$, and $Y$ is a pseudo-event full specification of $\cY$.  To apply QSO1 to these pairs we need to establish that the members of each pair occur in spacelike regions, and find the mutual past of those regions. Recall that this full specification $X \in \Phi_p(\cX)$ is defined so that $\pdom(X) \subset \cX$, and since $\cA$ is disjoint to $\cX$, $\pdom(A \cap X) \subset \cA \cup \cX$, for all $X\in \Phi_p(\cX)$, from lemma \ref{pseudoi} (similarly for $B$ and $Y$).  It is also the case that $\pdom(X) \subset \cA \cup \cX$ and $\pdom(Y) \subset \cB \cup \cY$.  From this we can see that for all four pairs of pseudo-events, the two pseudo-events occur in the pair of regions $\{(\cA \cup \cX),(\cB \cup \cY)\}$ respectively.  As noted before, it can be seen that $(\cA \cup \cX) \nat (\cB \cup \cY)$, and that the mutual past for this pair of regions is $\cP_1$.  Thus QSO1 can be applied four times, each time using a pseudo-event full specification $C$ of the past region $\cP_1$.  QSO1 therefore gives:

\begin{gather}
\forall X \in \Phi_p(\cX), Y \in \Phi_p(\cY), C \in \Phi_p(\cP_1) \notag \\
\hm(A \cap Y \cap C)\hm(B \cap X \cap C) = \hm(A \cap B \cap X \cap Y \cap C)\hm(C) \quad \text{, and} \\
\hm(A \cap Y \cap C)\hm(X \cap C) = \hm(A \cap X \cap Y \cap C)\hm(C) \quad \text{, and} \\
\hm(Y \cap C)\hm(B \cap X \cap C) = \hm(B \cap X \cap Y \cap C)\hm(C) \quad \text{, and} \\
\hm(Y \cap C)\hm(X \cap C) = \hm(X \cap Y \cap C) \hm(C).
\end{gather}
From lemma \ref{Alemd}, this implies that
\begin{gather}
\forall X \in \Phi_p(\cX), Y \in \Phi_p(\cY), C \in \Phi_p(\cP_1)  \notag \\
\label{Aprf1}
\hm(A \cap X \cap Y \cap C)\hm(B \cap X \cap Y \cap C)=\hm(A \cap B \cap X \cap Y \cap C)\hm(C) .
\end{gather}
From lemma \ref{pcora}, if $C \in \Phi_p(\cP_1)$, $X \in \Phi_p(\cX)$ and $Y \in \Phi_p(\cY)$, then $C \cap X \cap Y \in \Phi_p(\cP_2)$. Eqn. (\ref{Aprf1}) is therefore equivalent to (\ref{QSO2}) and so QSO2 is satisfied for the pair of pseudo-events $\{A,B\}$.
$\Box$
  
\begin{claim}\label{Aclaimb}
QSO2 implies QSO1.
\end{claim}

{\bf Proof:}  As above, take any pair of pseudo-events $A$ and $B$, such that $\pdom(A) \subset \cA$ and $\pdom(B) \subset \cB$ and $\cA \nat \cB$.  We only need note that, if $\pdom(A) \subset \cA$ and $\pdom(B) \subset \cB$ ,then it is also true that $\pdom(A) \subset \cA \cup \cX$ and $\pdom(B) \subset \cB \cup \cY$, where $\cX$ and $\cY$ are defined as before.  The joint past of the pair of regions $\{\cA \cup \cX,\cB \cup \cY \}$ is $\cP_1$.  Since $\cA \cup \cX \nat \cB \cup \cY$, QSO2 gives
\begin{equation}
 \hm(A \cap B|C) = \hm(A|C)\hm(B|C) \quad \forall C \in \Phi_p(\cP_1). \\
\end{equation}
From this, we know that for any pseudo-events $\pdom(A) \subset \cA$ and $\pdom(B) \subset \cB$, if $\cA \nat \cB$, then the above equation holds.  This is exactly QSO1.
$\Box$

\begin{corollary}
\label{Acorb}
QSO1 is equivalent to QSO2.
\end{corollary}

\bibliographystyle{apacite} 

\bibliography{causality3}

\end{document}